\documentclass{PoS}
\usepackage{xspace}
\usepackage{graphicx}
\usepackage{sidecap}
\let\jnl=\rmfamily
\def\refe@jnl#1{{\jnl#1}}%
\newcommand\aj{\refe@jnl{AJ}}%
\newcommand\actaa{\refe@jnl{Acta Astron.}}%
\newcommand\araa{\refe@jnl{ARA\&A}}%
\newcommand\apj{\refe@jnl{ApJ}}%
\newcommand\apjl{\refe@jnl{ApJ}}%
\newcommand\apjs{\refe@jnl{ApJS}}%
\newcommand\ao{\refe@jnl{Appl.~Opt.}}%
\newcommand\apss{\refe@jnl{Ap\&SS}}%
\newcommand\aap{\refe@jnl{A\&A}}%
\newcommand\aapr{\refe@jnl{A\&A~Rev.}}%
\newcommand\aaps{\refe@jnl{A\&AS}}%
\newcommand\azh{\refe@jnl{AZh}}%
\newcommand\memras{\refe@jnl{MmRAS}}%
\newcommand\mnras{\refe@jnl{MNRAS}}%
\newcommand\na{\refe@jnl{New A}}%
\newcommand\nar{\refe@jnl{New A Rev.}}%
\newcommand\pra{\refe@jnl{Phys.~Rev.~A}}%
\newcommand\prb{\refe@jnl{Phys.~Rev.~B}}%
\newcommand\prc{\refe@jnl{Phys.~Rev.~C}}%
\newcommand\prd{\refe@jnl{Phys.~Rev.~D}}%
\newcommand\pre{\refe@jnl{Phys.~Rev.~E}}%
\newcommand\prl{\refe@jnl{Phys.~Rev.~Lett.}}%
\newcommand\pasa{\refe@jnl{PASA}}%
\newcommand\pasp{\refe@jnl{PASP}}%
\newcommand\pasj{\refe@jnl{PASJ}}%
\newcommand\skytel{\refe@jnl{S\&T}}%
\newcommand\solphys{\refe@jnl{Sol.~Phys.}}%
\newcommand\sovast{\refe@jnl{Soviet~Ast.}}%
\newcommand\ssr{\refe@jnl{Space~Sci.~Rev.}}%
\newcommand\nat{\refe@jnl{Nature}}%
\newcommand\iaucirc{\refe@jnl{IAU~Circ.}}%
\newcommand\aplett{\refe@jnl{Astrophys.~Lett.}}%
\newcommand\apspr{\refe@jnl{Astrophys.~Space~Phys.~Res.}}%
\newcommand\nphysa{\refe@jnl{Nucl.~Phys.~A}}%
\newcommand\physrep{\refe@jnl{Phys.~Rep.}}%
\newcommand\procspie{\refe@jnl{Proc.~SPIE}}%
\newcommand{\Al}{$^{26}$Al\xspace}

\newcommand{\Ti}{$^{44}$Ti\xspace}

\newcommand{\Msol}{M\ensuremath{_\odot}\xspace}

\title{SPI Measurements of Nucleosynthesis Gamma-Rays}

\ShortTitle{SPI Nucleosynthesis Measurements}

\author{\speaker{Roland Diehl}$^{1,2}$,%\thanks{also: Munich Excellence Cluster "Origins and Evolution of the Universe"}}, 
{Frauke Alexander}$^{1,3}$,%\thanks{also: Universit\"atssternwarte, LMU M\"unchen, Germany}}, 
{Martin Krause}$^{1,2}$,%\thanks{also: Munich Excellence Cluster "Origins and Evolution of the Universe"}}, 
{Daniel Lubos}$^{1}$,  %\thanks{also: Munich Excellence Cluster "Origins and Evolution of the Universe"}\\
{Karsten Kretschmer}$^{4,1}$, 
{Pierrick Martin}$^{5,1}$, and  
{Wei Wang}$^{6,1}$\\
        $^1$Max Planck Institut f\"ur extraterrestrische Physik, D-85748 Garching, Germany\\
        $^2$Munich Excellence Cluster "Origins and Evolution of the Universe", D-85748 Garching, Germany\\
        $^3$Universit\"atssternwarte, Ludwig Maximilian Universit\"at, D-81679 M\"unchen, Germany\\
        $^4$Fran\c{c}ois Arago Centre, APC, Universit\'e Paris Diderot, F-75013 Paris, France\\
        $^5$IRAP, F-31028 Toulouse, France\\
        $^6$National Astronomical Observatories, CAS, Beijing 100012, China\\
        E-mail: \email{rod@mpe.mpg.de}}
\abstract{Studies based on the gamma-ray lines from radioactive decay of unstable isotopes produced in massive-star and supernova nucleosynthesis have been among INTEGRAL's prominent science achievements. $^{26}$Al has become a tool to study specific source regions, such as massive-star groups and associations in nearby regions which can be discriminated from the galactic-plane background, and the inner Galaxy where Doppler shifted lines add to the astronomical information. $^{60}$Fe is co-produced by the sources of $^{26}$Al, and the isotopic ratio from their nucleosynthesis encodes stellar-structure information. Here we summarize latest results using the accumulated multi-year database of observations, point to the relevant publications, and discuss their astrophysical implications.}

\FullConference{"An INTEGRAL view of the high-energy sky (the first 10 years)"
9th INTEGRAL Workshop and celebration of the 10th anniversary of the launch,\\
		October 15-19, 2012\\
		Bibliotheque Nationale de France, Paris, France}

\begin{document}

\section{SPI and Nucleosynthesis Lines}
Gamma-ray lines from radioactive decays of unstable isotopes from nucleosynthesis are among the main science themes of SPI \cite{Vedrenne:2003} on INTEGRAL \cite{Winkler:2003}. The lessons learned from such measurements have recently been summarized \cite{Diehl:2013}. Direct detections of such lines has been reported from supernovae SN1987A  \cite{Matz:1988} and Cas A \cite{Iyudin:1994}, while indirectly the diffuse interstellar glow of $^{26}$Al \cite{Mahoney:1982} has been associated to cumulative massive-star synthesis \cite{Prantzos:1995}. SPI has been a follow-up experiment upon those earlier discoveries, and exploits superior spectral resolution \cite{Roques:2003} to measure those lines in greater detail, and to discover more if possible.

SPI confirmed the interstellar glow throughout the Galaxy of $^{26}$Al gamma-rays early in the INTEGRAL mission, and has been refining those measurements since. Later,  gamma-rays from decay of $^{60}$Fe also were discovered from the plane of the Galaxy. This signal was found to be much fainter than the one from $^{26}$Al, although the same massive-star populations are the most-plausible sources for both those relatively long-lived isotopes which decay long after they have been injected into interstellar space from their sources. The diffuse glow of positron annihilation emission, which should go hand in hand with nucleosynthesis of isotopes on the proton-rich side of the stability valley of isotopes, such as $^{26}$Al or $^{44}$Ti or $^{56}$Ni, has not yet been detected unambiguously from the Galaxy as a whole, and also not from the prominent $^{26}$Al-bright massive-star regions.

\begin{SCfigure}%[th] %%%%%%%%%%%%%%%%%%%%%%%%%%%%%%%%%%
\centering
\includegraphics[width=0.7\textwidth]{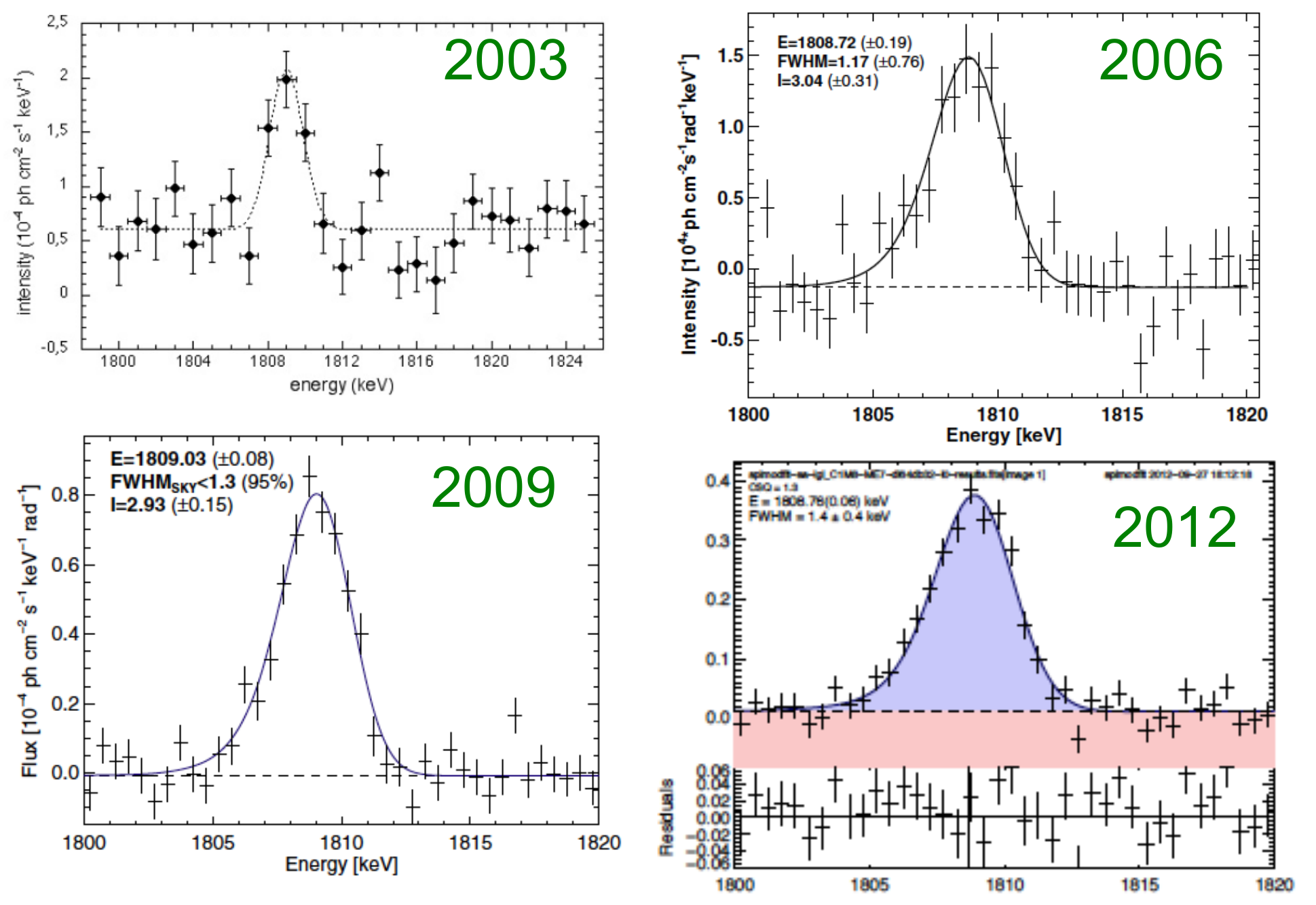}
\caption{The \Al line measurements with SPI during the INTEGRAL mission.}
\label{fig_Al-spectra_years}
\end{SCfigure}%%%%%%%%%%%%%%%%%%%%%%%%%%%%%%%%%%

Detecting individual nucleosynthesis sources through their radioactive afterglows remains a matter of luck: Sources must be sufficiently nearby to produce gamma-ray line fluxes above SPI's sensitivity of typically 3~10$^{-5}$~ph~cm$^{-2}$s$^{-1}$ (for a Ms exposure). Up to now, neither a supernova nor a nova clearly exceeded this detection threshold; see below for $^{44}$Ti from core-collapse events and $^{56}$Ni decay chain gamma-rays from SNIa. But predicted rates of sufficiently-nearby events are in the range of one per year, and therefore, with the uncertainty of such small-number statistics, it appears worthwhile to await SPI's chance in the extended INTEGRAL mission.  

\section{Diffuse nucleosynthesis lines: $^{26}$Al}
Measurements of \Al with SPI have been restricted to 'single events' up to now \cite{Diehl:2003a,Diehl:2006c,Wang:2009}. Detected events which trigger more than one detector, deposit different fractions of the total gamma-ray energy in these detectors, and the total energy deposit of a detected event would be the sum of different energies across the range from threshold up to the photo peak energy \cite{Vedrenne:2003}. SPI's 19 Ge detectors have some scatter in their individual spectral response functions and their energy dependence. Calibrations ensure that there is a close match of responses at the same energy (e.g. of the 1808.63~keV line from \Al), but responses might be less homogeneous across different energies. Selecting single-detector events only avoids that uncertainty, hence has been applied so far. But at those energies, only 1/2 of all detections are 'single events', the other 1/2 being multiple-detector hits which have been filtered out so far. Recently, we have compared spectral line shapes above 1~MeV for different lines among single and multiple events, and find that differences only become significant in the tails of the lines, well outside the central (FWHM) parts of the line. Hence, we performed new analyses also including multiple-detector events. The major challenge here is to also establish a reliable instrumental-background model for those multiple-detector events.

\begin{SCfigure}%[th] %%%%%%%%%%%%%%%%%%%%%%%%%%%%%%%%%%
\centering
\includegraphics[width=0.4\textwidth]{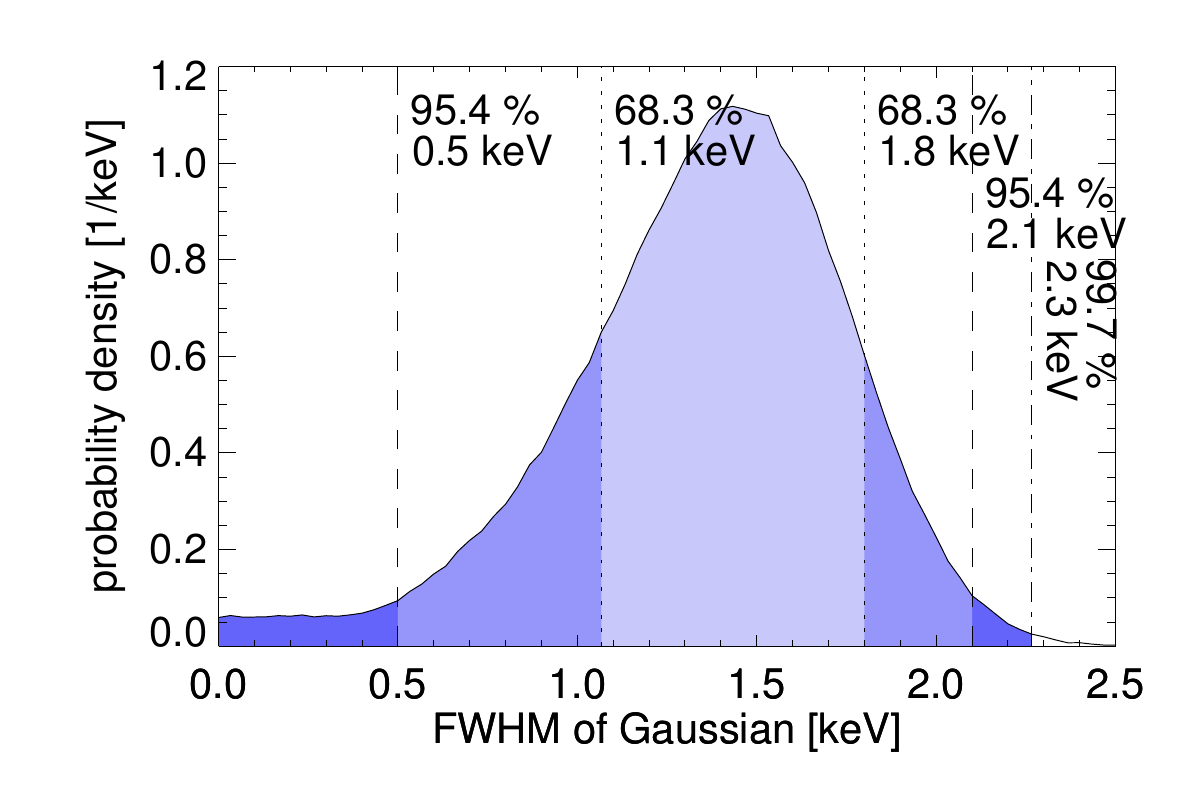}
\caption{The astrophysical  \Al line width constraint.}
\label{fig_Al-width_2012}
\end{SCfigure}%%%%%%%%%%%%%%%%%%%%%%%%%%%%%%%%%%

Figure~\ref{fig_Al-spectra_years} shows the measured \Al signal from the entire Galaxy, adopting spatial distribution on the sky according to COMPTEL measurements \cite{Pluschke:2001c}, as derived for different datasets over the years, the latest result using also multiple events and data up to orbit number 1142 (February 2012). The \Al line parameters of intensity, line position, and width, are all consistent with earlier results based on single events only. The statistical precision increases from the larger number of events, although some additional uncertainty from background modeling arises. Nevertheless, the \Al detection arises at 32~$\sigma$ significance, from those data.   
Figure~\ref{fig_Al-width_2012} shows the probability distribution of the line width for an additional Gaussian line broadening beyond the instrumental resolution alone. Here we report a first positive measurement of astrophysical line broadening, with a value of 1.4~keV ($\pm$0.3~keV), which corresponds to 175~km~s$^{-1}$ ($\pm$45~km~s$^{-1}$) in velocity space. For comparison, the expectations from averaging over different velocities from large-scale rotation of the Galaxy are about 1.35~keV, and compatible with this measurement.

\begin{SCfigure}%[th] %%%%%%%%%%%%%%%%%%%%%%%%%%%%%%%%%%
\centering
\includegraphics[width=0.6\textwidth]{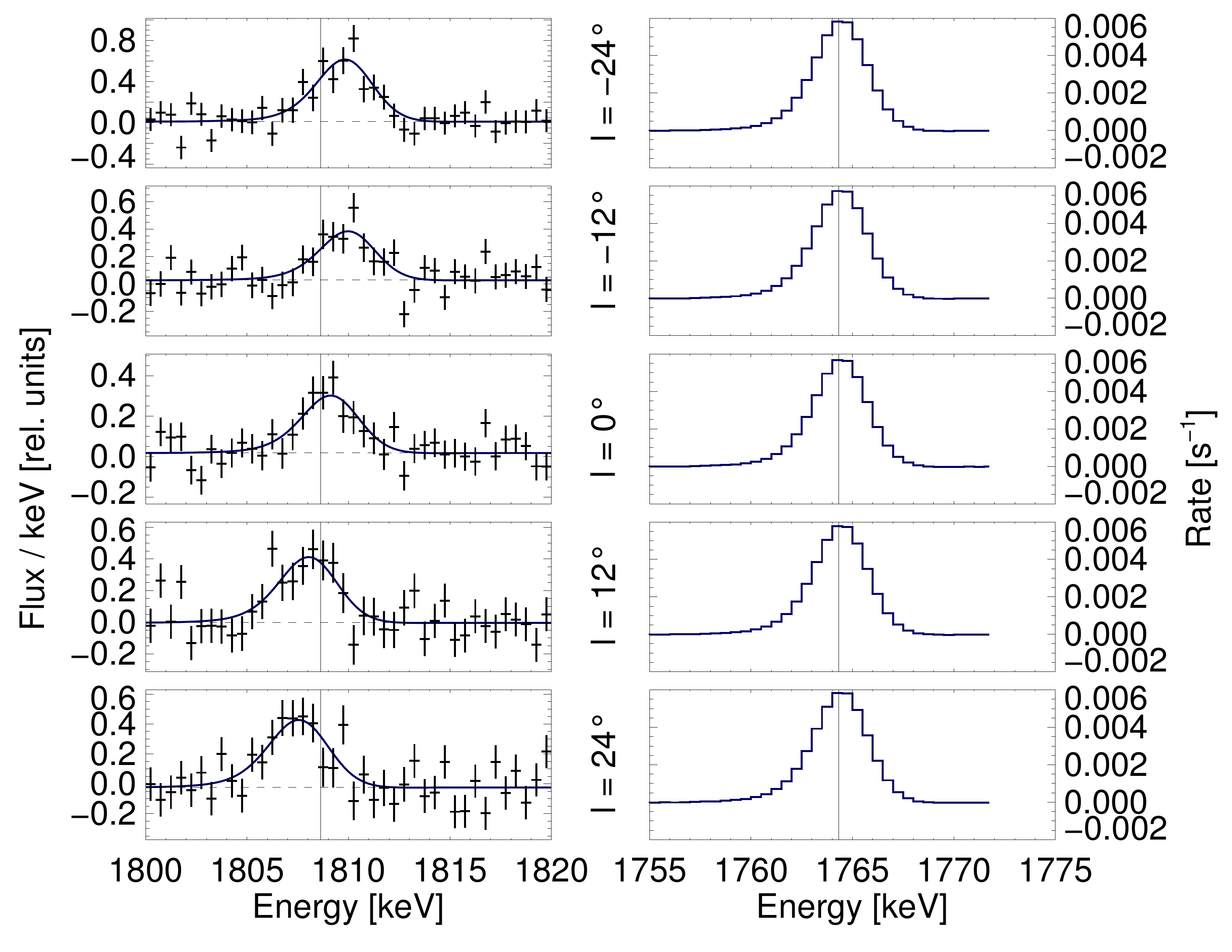}
\caption{The \Al line measurements with SPI along the plane or the Galaxy, for different lines of sight. The systematic Doppler shift from large-scale rotation in the galaxy can be seen clearly (\emph{left}). A nearby instrumental line at 1764 keV (\emph{right}) does not show differences in line position, from the same data.}
\label{fig_Al-spectra_plane}
\end{SCfigure}%%%%%%%%%%%%%%%%%%%%%%%%%%%%%%%%%%

The imprint of large-scale Galactic rotation on the measurements of \Al with SPI had been reported earlier: Bulk motion differs for different lines of sight along the plane of the Galaxy, with systematic variations leading to a blue shift of the \Al line towards the 4$^{th}$ quadrant of the Galaxy, and a red shift towards the 1$^{st}$ quadrant, correspondingly. The Doppler shift signature now is much clearer, with additional data, and more-sophisticated spatial analysis (see Fig.\ref{fig_Al-spectra_plane}). Surprisingly, \Al enriched gas appears to show systematically larger velocities than expected. This suggests that the ejection of nucleosynthesis material from massive-star sources occurs into surrounding interstellar medium which has been shaped by massive-star winds and the gas accumulation in spiral arms in peculiar ways. These implications of \Al-measured Doppler shifts and their comparison to longitude-velocity measurements for other Galactic components are discussed elsewhere (Kretschmer et al. submitted 2012). 

The brightest individual \Al emission region appears to be the Cygnus region. From the nine OB associations that potentially contribute to the signal, probably Cyg OB2 dominates by far \cite{Martin:2010b}. It has been found that predicted yields for the stellar population as we know it for this region and their characteristic metalicity fall significantly below the intensity as observed by SPI. This could either mean that we underestimate (still) the richness of the stellar population, or else the \Al yields of the Wolf-Rayet wind contributors are higher than current models predict, because those sources should dominate over core-collapse supernovae for the ages of stars inferred for Cygnus. 

Similar studies have been made for the Carina \cite{Voss:2012} and Orion \cite{Voss:2010a} regions, where \Al signals have not been detected up to now with INTEGRAL. SPI exposure in Carina is substantial, and the non-detection actually sets an upper limit on wind-produced interstellar \Al here \cite{Voss:2012}. The SPI exposure in Orion is low, and an observing program has been accepted recently to change this. SPI should be able to measure \Al emission \cite{Voss:2010a} , following the $^{26}$Al detection in the COMPTEL sky survey \cite{Diehl:2003e}. The Orion region presents an interesting situation, as the \Al source (Orion OB1 stars) is located on the near side of the Orion molecular clouds, with likely preference to blow ejecta into the Eridanus cavity which extends from the Orion molecular clouds towards the Sun by about 300~pc. Attempts to determine the X-ray emission throughout the Eridanus cavity, thus to determine if interiors are heated by nucleosynthesis events or rather shock heating at the walls dominates, have not been successful with XMM-Newton so far \cite{Lubos:2012}. Early ROSAT results were promising \cite{Snowden:1995}, but a dedicated XMM-Newton observing strategy is required for this extended sky region.    

The nearby region corresponding to stars from the Scorpius-Centaurus association has been discriminated against the Galactic emission in \Al \cite{Diehl:2010}. Similar to the Orion region, this promises that even modest spatial telescope resolution such as from SPI (2.7$^{\circ}$) could reveal \Al emission displaced from its sources, and thus teach us about ejecta flows. The \Al ejected from the Scorpius-Centaurus stars should in fact be distributed over a large region on the sky. But the distribution might help to distinguish contributions from different-age subgroups, thus helping to study the scenario of triggered star formation (see presentation by Alexander et al., this conference). 

%%%%%%%%%%%%%%%%%%%%%%%%%%%%%%%%%

\begin{SCfigure}%[th] %%%%%%%%%%%%%%%%%%%%%%%%%%%%%%%%%%
\centering
\includegraphics[width=0.5\textwidth]{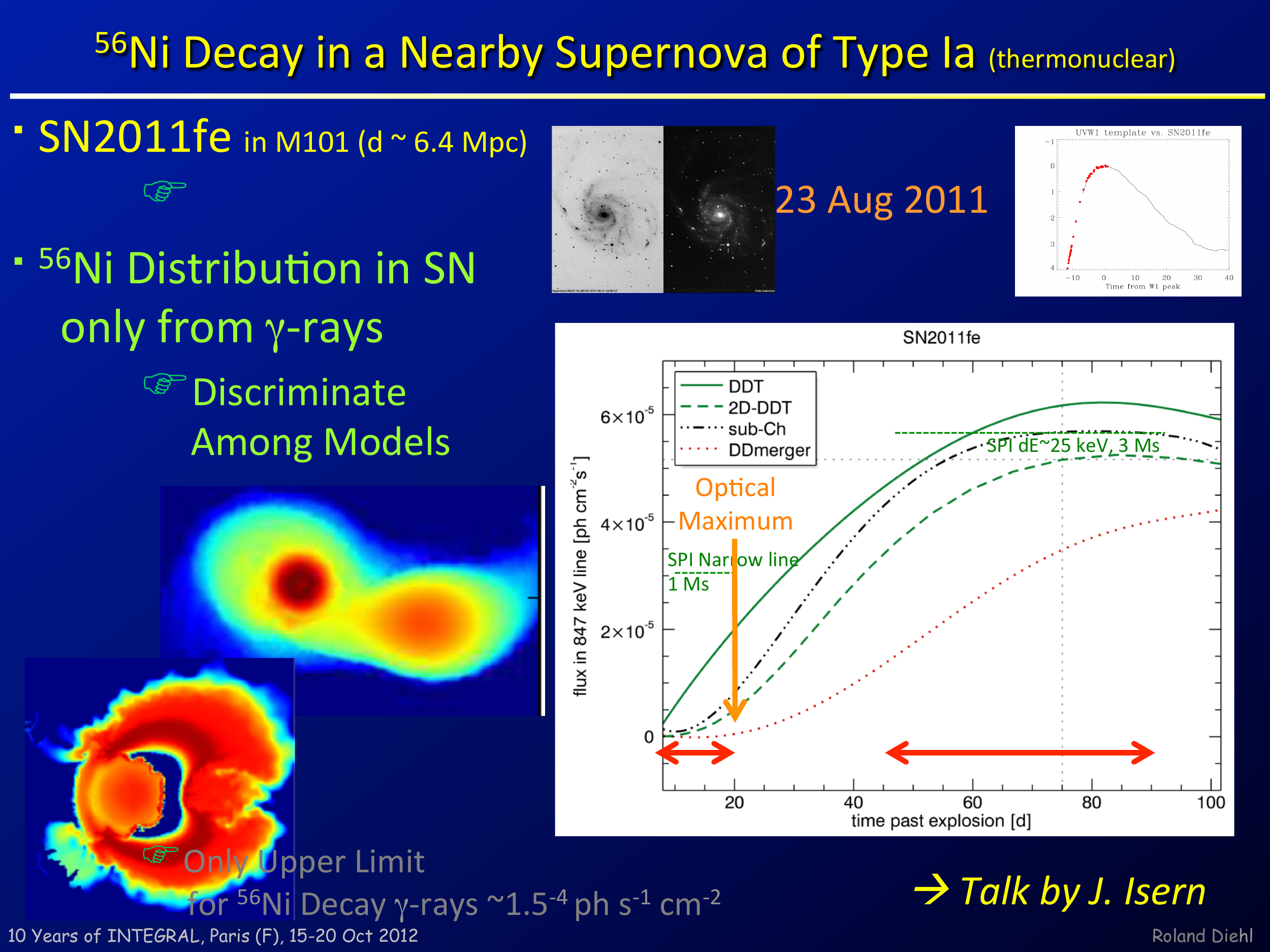}
\caption{The supernova SN2011fe occurred in August 2011, and is the most-nearby Type Ia supernova that occurred during the INTEGRAL mission. The maximum brightness at optical wavelengths occurs about 20 days after the explosion (orange arrow). The brightness in gamma-rays from the $^{56}$Ni decay chain is shown to evolve differently with time for different SNIa models. INTEGRAL observing times are indicated (red arrows) and sum up to 4 Ms. SPI measurements were not sensitive enough for a detection, however \cite{Isern:2011,Isern:2011b}. }
\label{fig_SN2011fe}
\end{SCfigure}%%%%%%%%%%%%%%%%%%%%%%%%%%%%%%%%%%

\section{Supernovae}
Supernova light is powered from radioactive $^{56}$Ni, ejected in substantial amounts (0.1 to 0.7 \Msol) during these explosions. The amounts released depend on the physical mechanism and morphology of the explosion. Generally, supernovae should become transparent to gamma-rays from radioactive decays within the first weeks to months, and thus gamma-ray brightness evolution during these phases will reveal how deeply embedded the $^{56}$Ni had been. For core-collapse events, \Ti is also produced at significant amounts inside the supernova, but its ejection depends critically on the location of the mass cut which separates the ejecta from matter ending up in the compact remnant star. 

Supernova SN2011fe \cite{Nugent:2011} occurred on 24 August 2011, and is the most-nearby Type Ia supernova that occurred during the INTEGRAL mission. Its host galaxy M101  is reported to have a distance of 6.4~Mpc.  Figure~\ref{fig_SN2011fe} shows how the gamma-ray brightness is expected to evolve according to currently-favored model classes: a delayed detonation (DDT, shown for a 1- and 2-dimensional treatment), sub-Chandrasekhar, and (double-degenerate) white-dwarf merger models. Also indicated are the optical maximum brightness, and the two INTEGRAL observing periods, scheduled to try and exploit this unique opportunity.  But data analysis so far did not show any of the candidate lines from the $^{56}$Ni decay chain in the early observations \cite{Isern:2011,Isern:2011b}. 

\begin{SCfigure}%[th] %%%%%%%%%%%%%%%%%%%%%%%%%%%%%%%%%%
\centering
\includegraphics[width=0.6\textwidth]{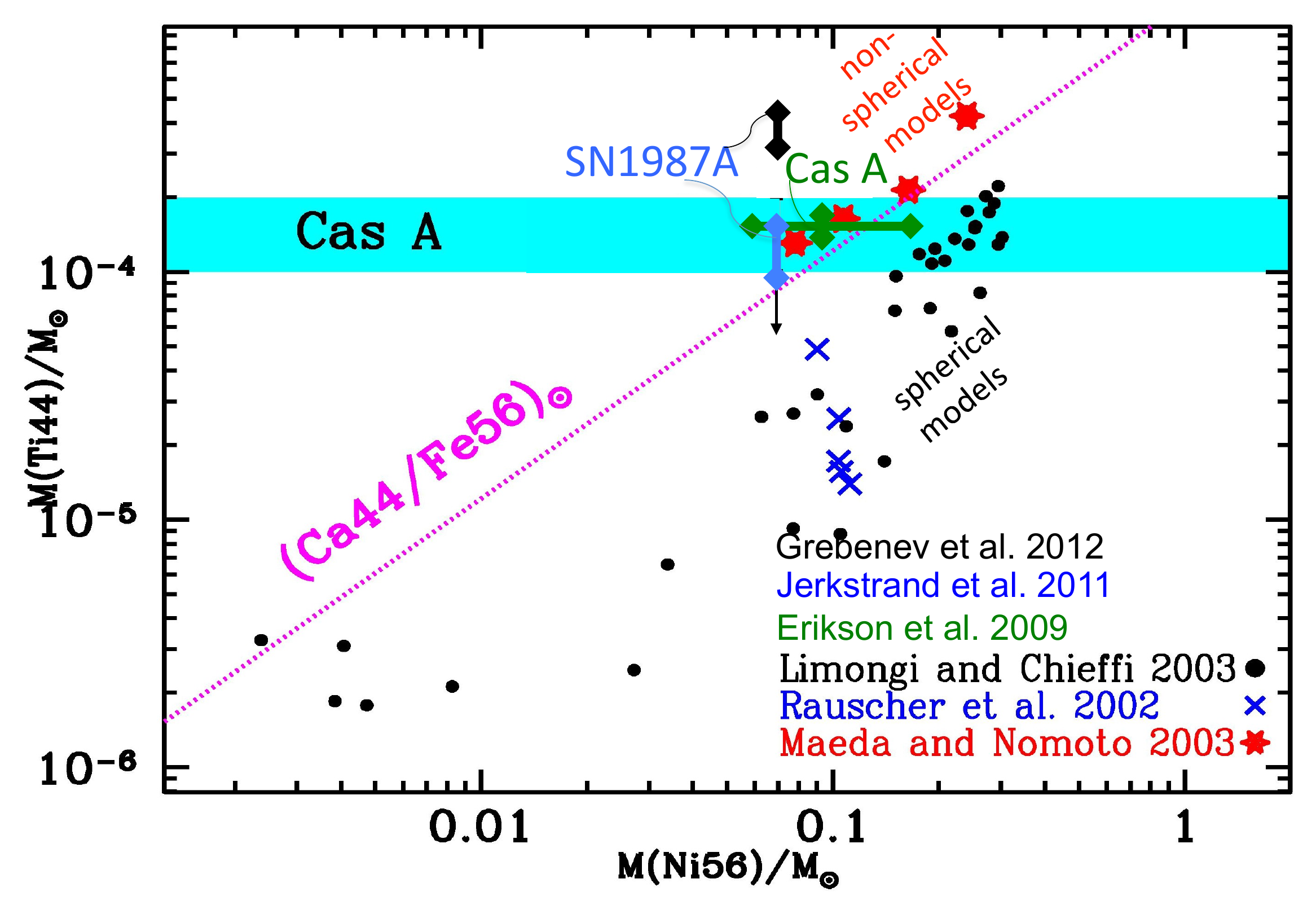}
\caption{The ratio of radioactive $^{56}$Ni to \Ti for different models (dots=spherical, crosses=aspherical are compared with constraints for the standard abundances (line), and for the two prominent core-collapse events of SN1987A and Cas A. (from \cite{Diehl:2013})}
\label{fig_Ni-Ti-ratio}
\end{SCfigure}

The ratio of radioactive $^{56}$Ni to \Ti is a sensitive probe of how material from the inner core-collapse region of a supernova may be ejected \cite{Diehl:1998}. Models for spherical explosions generally predict rather little \Ti to be ejected, while aspherical explosions may boost ejected \Ti amounts. The two prominent core-collapse events of SN1987A \cite{Grebenev:2012} and Cas A \cite{Renaud:2006a,Martin:2009a} have now been measured with gamma-ray telescopes in \Ti gamma-rays, and $^{56}$Ni amounts have been inferred from other observations for each of those (though indirectly for Cas A's explosion dated to about 1668). The ratio of $^{44}$Ti to $^{56}$Ni can thus be determined, although uncertainties are large (typically 30\%), and compared to the ratio inferred from solar abundances of daughter isotopes, as well as to those model families (see Fig.~\ref{fig_Ni-Ti-ratio}). It turns out that non-spherical explosions appear necessary to explain the observed events, quite plausible in view of the non-spherical phenomena observed for each of these events. Apparently, \Ti ejection is not a common outcome of most of the core-collapse events (see \cite{The:2006} for a detailed discussion), as also supported from the absence of more \Ti sources in Galactic-plane surveys (see Renaud et al., these Proceedings). 

%%%%%%%%%%%%%%%%%%%%%%%%%%%%%%%%%%

\section{Summary}
SPI continues to refine the signal from nucleosynthesis gamma-rays in several key areas. $^{26}$Al measurements now are above thresholds to split the total-Galaxy signal into spatially-resolved analyses, such as individual massive-star regions, or trends along the plane of the Galaxy. Studies of individual massive-star regions have indicated that the yields in $^{26}$Al for massive stars may need to be revised, hinting at effects of winds during the Wolf-Rayet phase and their detailed properties. The global $^{60}$Fe/$^{26}$Al ratio demonstrates that current models of massive-star structure in the late pre-supernova phase may be incorrect, as $^{60}$Fe yields are predicted towards the high side of observational constraints. Yet, since this constraint is derived for an entire population of sources, possibly the distribution of stellar-masses in the stellar population could differ from expectations, as an alternative explanation. Measurements from supernovae are sparse, in contrast. Yet, $^{44}$Ti measurements from young supernova remnants Cas A and SN1987A remain exciting, as those two objects largely determine our detailed knowledge about core-collapse physics. SN1987A appears beyond SPI's reach, while for Cas A the non-detection, combined with SPI's spectral resolution and background properties, constrains $^{44}$Ti nuclei to still move at velocities above 500 km~s$^{-1}$ in the 360-year-old remnant. The search for the brightest $^{56}$Ni decay chain lines from supernovae of type Ia still is ongoing, and SN2011fe at about 6 Mpc distance still was too far away, as supernova lines are kinematically broadened. SPI is the last nuclear-line telescope for years to come. Therefore, continuing these types of measurements will be worthwhile, both for refinements of the $^{26}$Al, $^{60}$Fe, and positron annihilation results, and for eventually catching data from a nearby explosive event.  

\noindent{\bf Acknowledgements.}
The INTEGRAL/SPI project has been completed under the responsibility and leadership of CNES; we are grateful to ASI, CEA, CNES, DLR (grants 50 OG 1101 and 50 OR 0901), ESA, INTA, NASA and OSTC for support of this ESA space science mission and its science analysis. 
This research was supported by the German DFG cluster of excellence \emph{Origin and Structure of the Universe}, and by the DFG priority program \emph{Physics of the ISM} (grant PR569/10-1). 

%%%%%%%%%%%%%%%%%%%%%%%%%%%%%%%%%%%%%%%%%%%%%%%%%%%%%%%%%%
%\bibliographystyle{harvard} 
\bibliographystyle{abbrv}%{aipproc}%{apalike}%{abbrv} 

\end{document}